\documentclass[%
 reprint,
superscriptaddress,
amsmath,amssymb,
aps,
pra,
]{revtex4-2}
\usepackage{graphicx}% Include figure files
\usepackage{dcolumn}% Align table columns on decimal point
\usepackage{bm}% bold math
\usepackage{float}
\usepackage{multirow}
\usepackage[colorlinks,linkcolor=blue, urlcolor=blue, anchorcolor=blue, citecolor=blue]{hyperref}
\begin{document}

%\preprint{APS/123-QED}

\title{Quantum battery optimized by parametric amplification}% Force line breaks with \\
\author{Fang-Mei Yang}%
\affiliation{College of Physics and Electronic Engineering, Northwest Normal University, Lanzhou, 730070, China}
\affiliation{Gansu Provincial Research Center for Basic Disciplines of Quantum Physics, Lanzhou, 730000, China}
\author{Jun-Hong An}
\affiliation{Key Laboratory of Quantum Theory and Applications of MoE, Lanzhou Center for Theoretical Physics, and Key Laboratory of Theoretical Physics of Gansu Province, Lanzhou University, Lanzhou 730000, China}
\affiliation{Gansu Provincial Research Center for Basic Disciplines of Quantum Physics, Lanzhou, 730000, China}
\author{Fu-Quan Dou}
\email{doufq@nwnu.edu.cn}
\affiliation{College of Physics and Electronic Engineering, Northwest Normal University, Lanzhou, 730070, China}
\affiliation{Gansu Provincial Research Center for Basic Disciplines of Quantum Physics, Lanzhou, 730000, China}

\begin{abstract}
The parametric amplification enabled by two-photon driving constitutes a versatile platform for advanced quantum technologies. We present an optimized scheme for implementing quantum batteries (QBs) based on a superconducting circuit system, where a two-photon-driven LC resonator serves as the charger and an array of transmon qubits functions as the battery. Our results show that two-photon parametric driving exponentially enhances the effective cavity-qubit coupling, which in turn gives rise to near-degenerate energy-level structures and highly entangled quantum states. This significantly enhances the charging power and enables rapid energy transfer from the charger to the battery. Moreover, the engineered squeezed cavity mode and the associated quantum correlations effectively suppress environmentally induced decoherence, thereby delaying energy leakage and facilitating stable energy storage. The proposed scheme remains robust against practical experimental imperfections, such as parameter disorder and environmental noise, preserving its performance advantages. The work provides a feasible platform for realizing high-power, high-stability QBs and highlights the potential of parametric control in quantum energy technologies.
\end{abstract}
\maketitle
%\tableofcontents
\section{Introduction}
Quantum batteries (QBs) are emerging energy storage devices based on nonequilibrium quantum thermodynamics, consisting of quantum systems with discrete energy levels that are excited to temporarily store energy \cite{PhysRevE.87.042123,RevModPhys.96.031001,Ferraro2026}. Their core mechanism relies on multipartite quantum correlations to realize superlinear scaling in charging power, thereby exhibiting a quantum advantage over classical linear scaling \cite{PhysRevB.99.205437,PhysRevLett.128.140501,PhysRevLett.125.236402,PhysRevLett.127.100601,PhysRevLett.118.150601,PhysRevB.105.115405,PhysRevA.103.052220,PhysRevA.105.062203,PhysRevA.106.032212,PhysRevE.103.042118}. This advantage has motivated extensive studies, mainly devoted to model innovation \cite{PhysRevA.97.022106,PhysRevA.104.032606,PhysRevE.100.032107,PhysRevA.110.032205,PhysRevB.109.235432,Shaghaghi2022,PhysRevLett.132.210402,PhysRevResearch.4.013172,8xsm-5mb6,l39v-jwwz,PhysRevLett.134.180401}, performance optimization \cite{PhysRevA.100.043833,PhysRevLett.124.130601,PhysRevA.107.023725,PhysRevA.102.052223,PhysRevLett.132.090401,PhysRevLett.122.210601,PhysRevA.109.062432,PhysRevA.103.033715,PhysRevLett.131.060402,PhysRevLett.133.197001,Sun2025,PhysRevLett.134.010408,PhysRevLett.134.220402,stc3-xkp5,PhysRevA.102.060201}, and experimental validation \cite{quach2020,PhysRevA.106.042601,PhysRevLett.131.260401,PhysRevA.107.L030201,Hu2022,Zheng2022,PhysRevLett.131.240401,sp5l-c6m8,y3qx-cs3r}. The coupling strength between the battery and the charger directly determines the charging power and serves as the key tuning parameter for realizing quantum advantage \cite{RevModPhys.96.031001,PhysRevLett.120.117702,RevModPhys.91.025005}.
In practice, overly weak coupling restricts the charging rate, whereas excessively strong coupling both exacerbates decoherence and poses experimental difficulties. Therefore, optimizing the trade-off between coupling strength and decoherence, together with developing precise control techniques, constitutes a critical challenge for current experimental research on QBs.

The strong coupling regime is essential for observing coherent quantum dynamics between light and matter \cite{RevModPhys.91.025005}. Significant progress in nanofabrication and quantum control has enabled the realization of strong coupling across diverse platforms, including superconducting circuits, semiconductor quantum dots, and solid-state spin systems \cite{PhysRevLett.96.057405,Wallraff2004,PhysRevX.7.011030,PhysRevLett.105.140501,PhysRevLett.105.140502,Petersson2012}. In 2016, two independent experiments attained a qualitative jump in the light-matter interaction strength, pushing the boundaries into the nonperturbative ultra-strong coupling domain by using Josephson junctions as coupling elements \cite{Yoshihara2017,Forn2017,Gheeraert2017,PhysRevX.2.021007,Gunter2009,Niemczyk2010}. Subsequently, the deep-strong coupling regime has been experimentally realized in both superconducting circuits and in two-dimensional electron gases coupled to terahertz metamaterial resonators \cite{Langford2017,PhysRevLett.105.263603}. In two regimes, where light-matter interactions approach or exceed the bare frequencies of the uncoupled systems, counter-rotating terms in the Hamiltonian become nonnegligible, leading to novel dynamics and entanglement properties \cite{PhysRevA.96.013849,PhysRevB.79.201303,PhysRevLett.105.237001,PhysRevLett.107.190402,PhysRevLett.108.120501,PhysRevA.94.012328,PhysRevLett.119.183602}. Nevertheless, current implementations are generally limited by fabrication complexity, constrained tunability, and scalability challenges, which hinder the simultaneous attainment of high coupling strengths, precise dynamic control, and large-scale integration. Thus, developing generalizable approaches that do not depend on highly specialized materials or device architectures is crucial for unlocking the full potential of these extreme coupling regimes in future quantum applications.

Parametric amplification techniques, especially those based on nonlinear two-photon driving, provide a promising physical pathway to address these limitations \cite{RevModPhys.93.041002,RevModPhys.84.1,PhysRevLett.114.093602,PhysRevLett.93.207002,Lemonde2016,PhysRevLett.128.183602,PhysRevLett.125.203601,PhysRevLett.120.093602}. These approaches utilize intrinsic nonlinearities, such as Josephson or Kerr nonlinearity, to enable frequency conversion and mode coupling, thereby inducing effective ultra-strong or deep-strong coupling dynamics in systems that are otherwise strongly coupled \cite{PhysRevLett.120.093602,PhysRevA.108.013512,PhysRevA.109.062423,PhysRevA.94.033841}. This strategy not only eliminates the dependence on specialized materials or device architectures, but also allows dynamic tuning of coupling strengths through control parameters including drive amplitude and frequency. Furthermore, two-photon parametric processes naturally generate squeezed states, which can further suppress environmental noise and enhance system coherence \cite{PhysRevLett.120.093602,PhysRevX.6.031004,PhysRevLett.113.220502}. Consequently, parametric amplification via two-photon driving offers new theoretical and experimental prospects for optimizing and realizing quantum systems in the ultra-strong and deep-strong coupling regimes under flexible and controllable conditions.

In this work, we propose a superconducting quantum circuit consisting of a two-photon-driven LC resonator coupled to an array of transmon qubits, providing a physical platform for the optimization and realization of QBs. Our study focuses on the exponential enhancement of the cavity-qubit coupling via two-photon parametric amplification, elucidating its pivotal role in boosting the charging and storage properties of QBs. We explore rapid energy transfer dynamics supported by near-degenerate energy-level structures, while analyze how squeezed cavity modes and correlated quantum states suppress energy leakage during stable energy storage. By investigating the effects of parameter disorder and environmental dissipation, we further demonstrate the robustness of this scheme under realistic experimental conditions, thereby establishing a theoretical framework and a feasible pathway toward readily scalable, high-performance quantum energy storage systems.

The paper is organized as follows. A superconducting quantum circuit system is presented in Sec.~\ref{section2}, where we derive its Hamiltonian and analyze the mechanism underlying the exponentially enhanced coupling induced by two-photon driving. In Sec.~\ref{section3}, we construct a QB model based on this system, discuss its collective charging advantage, and examine its robustness against parameter disorder and environmental dissipation. The main results are summarized in Sec.~\ref{section4}. The quantum criticality and quantum fluctuation of the system is addressed in Appendix~\ref{sectionA}. In Appendix~\ref{sectionB}, a more generalized quantum battery model is established and an approximate bound for its charging power is derived.
\begin{figure}[tbp]
\centering
\includegraphics[width=0.45\textwidth]{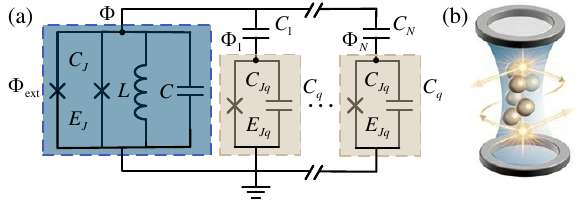}
\caption{(a) The schematic illustrates a superconducting circuit architecture consisting of an LC resonator, with inductance $L$ and capacitance $C$, capacitively coupled to an array of $N$ transmon qubits. Each qubit is modeled as a capacitor $C_q$ in parallel with a Josephson junction characterized by energy $E_{Jq}$ and capacitance $C_{Jq}$. The capacitive coupling is realized via the same capacitors $C_{j}$. A flux-pumped superconducting quantum interference device, formed by two Josephson junctions each with energy $E_J$ and capacitance $C_J$, is threaded by an external magnetic flux $\Phi_{\text{ext}}$ and embedded in the resonator loop to provide a tunable nonlinear inductive element. When the device is driven at twice the resonator frequency, it mediates an effective parametric two-photon driving to the cavity mode, enabling controlled squeezing of the microwave field. (b) Illustration of the QB. The golden arrows represent the optimization process facilitated by two-photon driving.}
\label{fig1}
\end{figure}
\section{Circuit QED} \label{section2}
We consider a superconducting quantum circuit comprising an LC resonator under two-photon driving and an ensemble of $N$ transmon qubits, shown in Fig. \ref{fig1}(a). The two-photon driving is implemented via a pair of Josephson junctions connected in parallel with the resonator, generating an effective squeezing Hamiltonian for the cavity mode. This design enables the controlled enhancement of quantum fluctuations and facilitates the exploration of strong correlations in cavity quantum electrodynamics at reduced physical coupling strengths \cite{PhysRevLett.93.207002}. The entire circuit is described by the following Lagrangian
\begin{equation*}
\mathcal{L}=\mathcal{L}_{cavity}+\mathcal{L}_{qubits}+\mathcal{L}_{coupling}+\mathcal{L}_{squeezing},
\end{equation*}
where
\begin{equation*}
\begin{aligned}
&\mathcal{L}_{cavity}=\frac{1}{2}C\dot{\Phi}^{2}-\frac{1}{2L}\Phi^{2},\\
&\mathcal{L}_{qubits}=\sum_{j=1}^{N}\left[\frac{1}{2}\left(C_{q}+C_{Jq}\right)\dot{\Phi_{j}}^{2}+E_{Jq}\cos\left(2\pi\frac{\Phi_{j}}{\Phi_{0}}\right)\right],\\
&\mathcal{L}_{coupling}=\sum_{j=1}^{N}\left[\frac{1}{2}C_{j}\left(\dot{\Phi}-\dot{\Phi_{j}}\right)^{2}\right],\\
&\mathcal{L}_{squeezing}=C_{J}\dot{\Phi}^{2}+2E_{J}\cos\left(\pi\frac{\Phi_{ext}}{\Phi_{0}}\right)\cos\left(2\pi\frac{\Phi}{\Phi_{0}}\right).
\end{aligned}
\end{equation*}
Here, the parameters $\Phi$ and $\Phi_{j}$ denote the magnetic fluxes at the respective nodes, $\Phi_{0}=h/2e$ is the flux quantum and $\Phi_{ext}=\Phi_{dc}+\delta\Phi\cos\left(\omega_{p}t\right)$ is the external flux with the parametric drive frequency $\omega_{p}$. The large Josephson energy restricts the phase to small fluctuations near zero, which justifies the expansion of the cosine terms in the Lagrangian to fourth order \cite{PhysRevA.76.042319}. The classical Hamiltonian is subsequently derived through the Legendre transformation
\begin{equation}
\begin{aligned}
H\approx&~\frac{1}{2C_{c}}Q^{2}+\frac{1}{2L_{c}}\Phi^{2}-\frac{\lambda_{0}}{\Phi_{0}^{2}}\cos\left(\omega_{p}t\right)\Phi^{2}\\
&+\sum_{j=1}^{N}\left[\frac{1}{2C_{q}}Q_{j}^{2}+\frac{1}{2L_{q}}\Phi_{j}^{2}-\frac{C_{j}}{C_{c}C_{q}}QQ_{j}\right],
\end{aligned}\label{clshmd}
\end{equation}
where $Q_{k}=\partial\mathcal{L}/\partial\Phi_{k}$ and $\lambda_{0}=4E_{J}\pi^{3}\frac{\delta\Phi}{\Phi_{0}}\sin(\pi\frac{\Phi_{dc}}{\Phi_{0}})$, with $k$ either empty or an index $j$.

In the weak-coupling regime, the cavity and the qubits can be quantized independently. Equation \eqref{clshmd} is thus quantized by requesting the canonical commutation relation $[\Phi_k,Q_k]=i\hbar$. The annihilation operators are
\begin{equation}
a=\frac{\Phi}{\sqrt{2\hbar Z_{c}}}+i\sqrt{\frac{Z_{c}}{2\hbar}}Q,~b_{j}=\frac{\Phi_{j}}{\sqrt{2\hbar Z_{q}}}+i\sqrt{\frac{Z_{q}}{2\hbar}}Q_{j}.
\end{equation}
From these relations, the quantized Hamiltonian is expressed as
\begin{equation}
\begin{aligned}
H=&~\hbar\omega_{c}a^{\dag}a-\frac{\hbar\lambda_{0}}{2C_{c}\Phi_{0}^{2}\omega_{c}}\cos\left(\omega_{p}t\right)\left(a+a^{\dag}\right)^{2} \\
&+\sum_{j=1}^{N}\left[\hbar\omega_{q}b^{\dag}_{j}b_{j}-\frac{\hbar C_{j}}{2\omega_{c}\omega_{q}}\left(ab_{j}^{\dag}+a^{\dag}b_{j}\right)\right],
\end{aligned}
\end{equation}
with the impedance $Z_{k_1}=\sqrt{L_{k_1}/C_{k_1}}$ and the frequency $\omega_{k_1}=1/\sqrt{L_{k_1}C_{k_1}}$ for $k_1=c,~q$. In a regime where the anharmonicity of the transmon is large, we can reduce the transmon to a two-level system \cite{PhysRevA.105.022423}. The Hamiltonian is truncated to two lowest levels and written as
\begin{equation}
\begin{aligned}
H=&~\hbar\omega_{c}a^{\dag}a-\hbar\lambda\cos\left(\omega_{p}t\right)\left(a+a^{\dag}\right)^{2}\\
&+\hbar\omega_{q}J_{z}-\hbar g\left(aJ_{+}+a^{\dag}J_{-}\right),
\end{aligned}\label{qhmt}
\end{equation}
the components of a collective spin operator are defined in terms of the Pauli operators as $J_{x,y,z}=\sum_{j=1}^{N}\sigma_{x,y,z}^{j}/2$. Here, $g=C_{j}/(\omega_{c}\omega_{q})$ and $\lambda=\lambda_{0}/(2C_{c}\Phi_{0}^{2}\omega_{c})$ denote the cavity-qubit coupling strength and the parametric drive amplitude, respectively.

We then move the time-dependent Hamiltonian \eqref{qhmt} into a frame rotating at half the parametric drive frequency $\omega_{p}$, with the unitary operator chosen as $U_{p}={\rm e}^{-i\omega_{p}t(a^{\dag}a+J_z)/2}$. The Hamiltonian becomes
\begin{equation}
H_{p}=\hbar\delta_{c}a^{\dag}a+\hbar\delta_{q}J_{z}-\hbar g\left(aJ_{+}+a^{\dag}J_{-}\right)-\frac{\hbar\lambda}{2}\left(a^{2}+a^{\dag2}\right),\label{hmd}
\end{equation}
where the cavity and qubit detunings are given by $\delta_{c/q}=\omega_{c/q}-\omega_{p}/2$. By a unitary transformation $U_{s}={\rm e}^{r/2\left(a^{\dag2}-a^{2}\right)}$, with a squeezing parameter $r$ defined as $\tanh\left(2r\right)=\lambda/\delta_{c}$, Eq. \eqref{hmd} is recast into
\begin{equation}
\begin{aligned}
H_{s}=&~\hbar\delta_{c}\operatorname{sech}{\left(2r\right)}a^{\dag}a+\hbar\delta_{q}J_{z}-\frac{\hbar g}{2}{\rm e}^{r}\left(a^{\dag}+a\right)\\
&\times \left(J_{+}+J_{-}\right)+\frac{\hbar g}{2}{\rm e}^{-r}\left(a^{\dag}-a\right)\left(J_{+}-J_{-}\right).\label{squeezedHamiltonian}
\end{aligned}
\end{equation}
The Hamiltonian takes the form of a generalized Dicke model with an additional term whose coupling strength decreases exponentially. Under strong parametric driving, the system becomes intrinsically equivalent to the standard Dicke model, characterized by an effective cavity detuning $\tilde{\delta}_{c}=\delta_{c}\operatorname{sech}(2r)$ and an enhanced coupling strength $\tilde{g}=ge^{r}/2$. The resulting effective critical coupling is given by $g_{c}=\sqrt{\delta_{c}\delta_{q}\operatorname{sech}(2r)}/2$, see Appendix~\ref{sectionA} for the derivation. The two-photon driving exponentially amplifies the cavity mode fluctuations, simultaneously increasing the effective light-matter coupling and reducing the effective critical coupling. Thus, the quantum phase transition occurs at a value significantly lower than the bare coupling strength. 
\begin{figure}[tbp]
\centering
\includegraphics[width=0.45\textwidth]{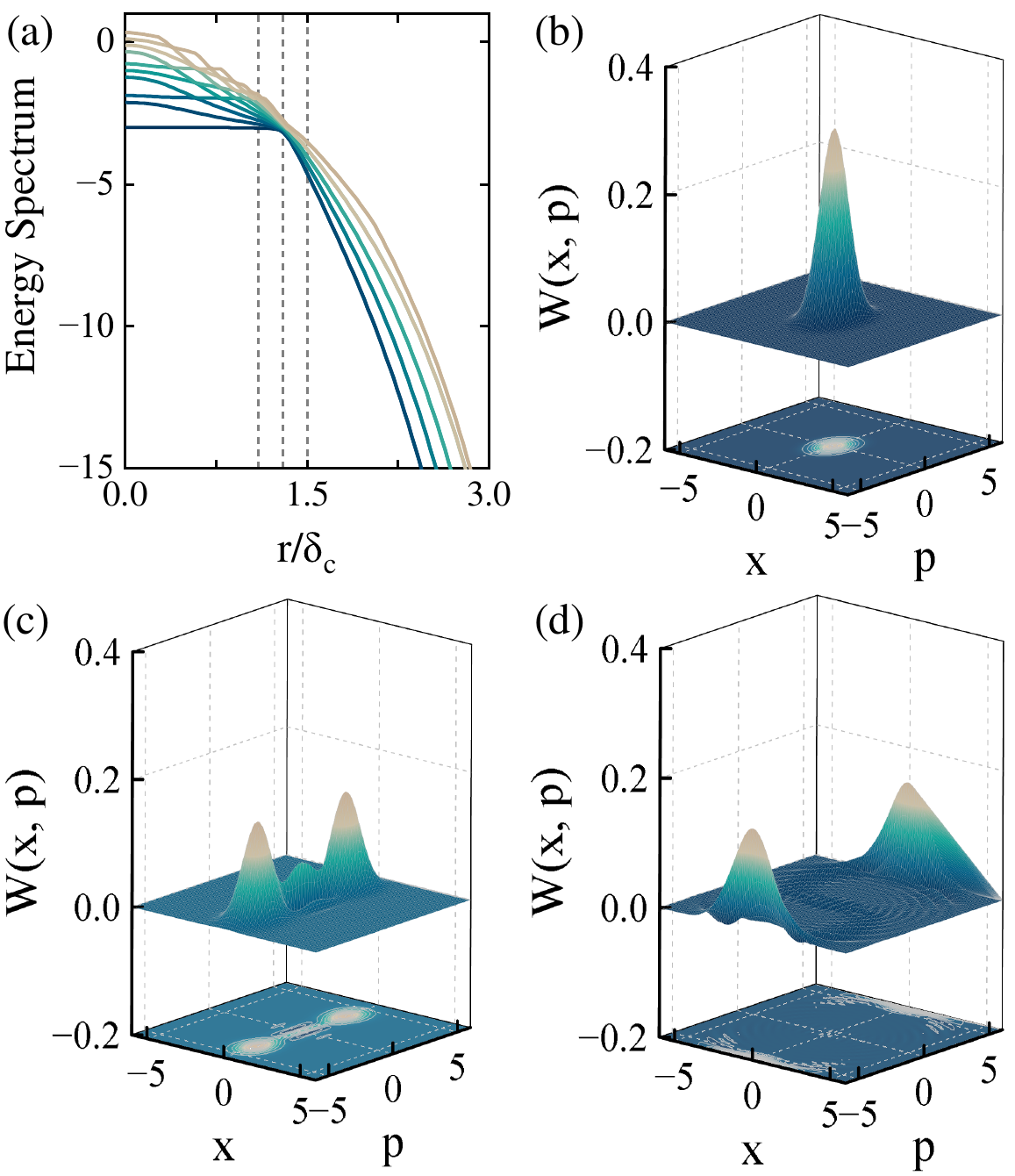}
\caption{(a) The first ten energy levels versus the squeezing parameter, with the number of qubits fixed at $N=6$. (b-d) The ground-state Wigner functions at different squeezing parameters $r/\delta_{c}=1.1$, $1.3$, $1.5$. Other parameters are $\delta_{q}/\delta_{c}=1$, and $g/\delta_{c}=0.05$.}
\label{fig2}
\end{figure}

We take into account the resonance case $\delta_{q}=\delta_{c}$. We find that the two-photon driving can shift the quantum phase transition threshold from the ultra-strong coupling regime $g/\delta_{c}=0.5$ without driving to the strong coupling regime $g/\delta_{c}=0.05$ at a squeezing parameter $r/\delta_{c}\approx1.22$. Figure \ref{fig2}(a) shows the variation of the energy spectrum with the squeezing parameter. Increasing this parameter drives the system from a non-degenerate strong-coupling regime to an ultra-strong-coupling regime with pairs of near-degenerate energy-level structures, concurrent with the normal-to-superradiant phase transition. In Figs.~\ref{fig2}(b)-\ref{fig2}(c), we introduce the Wigner quasi-probability distribution $W(x,p)=\frac{1}{\pi\hbar}\int_{-\infty}^{+\infty}\langle x+y|\rho_{g}|x-y\rangle{\rm e}^{-2ipy/\hbar}dy$ to probe quantum features, where $\rho_{g}$ is the ground-state density matrix of the cavity mode \cite{PhysRevLett.89.200402}. Across the normal-to-superradiant phase transition, the ground state evolves from a separable vacuum state into a quantum state with nonclassical entanglement. The corresponding Wigner function exhibits a pronounced bimodal structure with negative-value regions, indicating that significant quantum correlations and squeezing effects emerge during the quantum phase transition. Overall, parametric amplification enabled by two-photon driving provides enhanced coupling strength and macroscopic entangled states, forming a versatile platform for advanced quantum technologies. The resulting squeezed cavity field facilitates quantum metrology beyond the standard quantum limit \cite{PhysRevLett.117.030801,PhysRevA.73.033819}, while the exponentially amplified coupling underpins strongly correlated states that are indispensable for high-power, high-fidelity quantum devices, such as QBs.
\section{QB model and its charging advantages} \label{section3}
As illustrated in Fig. \ref{fig1}(b), our model features a battery $H_{b}=\hbar\omega_{q}J_{z}$ consisting of $N$ transmon qubits and a charger $H_{c}=\hbar\omega_{c}a^{\dag}a$ realized by a superconducting cavity. Moving beyond idealizations, real-world devices are inherently open systems susceptible to environmental noise. Therefore, the charging performance is quantified here under two dominant experimental decoherence channels. The charging dynamics is described by the following Lindblad master equation $\frac{d\rho}{dt}=-\frac{i}{\hbar}[H, \rho]+(\kappa\mathbb{L}_{a,a^\dag}+\gamma\mathbb{L}_{J_{-},J_+})\rho$, where the superoperator is defined as $\mathbb{L}_{o,p}\cdot=o\cdot p-\{ p o,\cdot\}/2$, $\kappa$ is the photon-loss rate from the cavity, and $\gamma$ is the collective relaxation rate of the qubit ensemble. Under the squeezed transformation $\rho_{s}=U_{s}^{\dagger}\rho U_{s}$, the dynamical equation retains the form
\begin{equation}
\frac{d\rho_{s}}{dt}=-\frac{i}{\hbar}[H_{s}, \rho_{s}]+(\kappa\mathbb{L}_{a_{s},a_s^\dag}+\gamma\mathbb{L}_{J_{-},J_+})\rho_{s},
\end{equation}
where the annihilation operator becomes $a_{s}=U_{s}^{\dagger}aU_{s}=a\cosh r+a^{\dagger}\sinh r$. The explicit form of $\mathbb{L}_{a_{s},a_s^\dag}\rho_s$ as
\begin{equation}
\begin{aligned}
\label{superoperator}
\mathbb{L}_{a_{s},a_s^\dag}\rho_{s}=&~[\cosh^2r\mathbb{L}_{a,a^\dag}+\sinh^2r\mathbb{L}_{a^{\dagger},a}\\
&+\cosh r\sinh r (\mathbb{L}_{a,a}+\mathbb{L}_{a^\dag,a^\dag})]\rho_s,
\end{aligned}
\end{equation}
reveals that in the squeezed frame, the single-photon loss term from the cavity is amplified, while additional gain-type and two-photon correlated dissipators emerge. This redistribution of environmental noise across phase space enables dynamical squeezing: quantum fluctuations are suppressed in one quadrature and amplified in the orthogonal one \cite{Krantz2019,3mlc-r8x4}. Thus, the two-photon driving effectively reshapes dissipation to facilitate noise suppression and stabilize nonclassical states.

The system is initialized with the cavity in an $n$-photon Fock state and the qubits in their collective ground state. The total initial state is a product state given by
\begin{equation}
|\Psi(0)\rangle=|n\rangle\otimes|\underbrace{G,\ldots, G}_{N}\rangle=|n; N/2, -N/2\rangle.
\end{equation}
The numerical solution of the Dicke model requires imposing a finite photon-number cutoff $n_{ph}$, since photon number is not conserved and the Hilbert space is unbounded in principle. In simulations, we set the initial photon number to $n=2N$ and the cutoff to $n_{ph}=4N+1$, which provides numerically stable, scalable results consistent with established methods \cite{PhysRevLett.120.117702,PhysRevB.102.245407,PhysRevB.105.115405}.

The charging performance is quantified by two figures of merit: the energy stored in the battery
\begin{equation}
E(t)={\rm Tr}[H_{b}\rho_{s}(t)]-{\rm Tr}[H_{b}\rho_{s}(0)],
\end{equation}
and the average charging power
\begin{equation}
P(t)=E(t)/t.
\end{equation}
Additionally, the maximal energy in the closed system, the stable energy in the open system, and the maximal power attainable at any time are respectively defined as
\begin{equation*}
E_{\rm max}={\rm max}(E),~E_{\rm sta}=E(\infty),~P_{\rm max}={\rm max}(P).
\end{equation*}

\begin{figure}[tbp]
\centering
\includegraphics[width=0.45\textwidth]{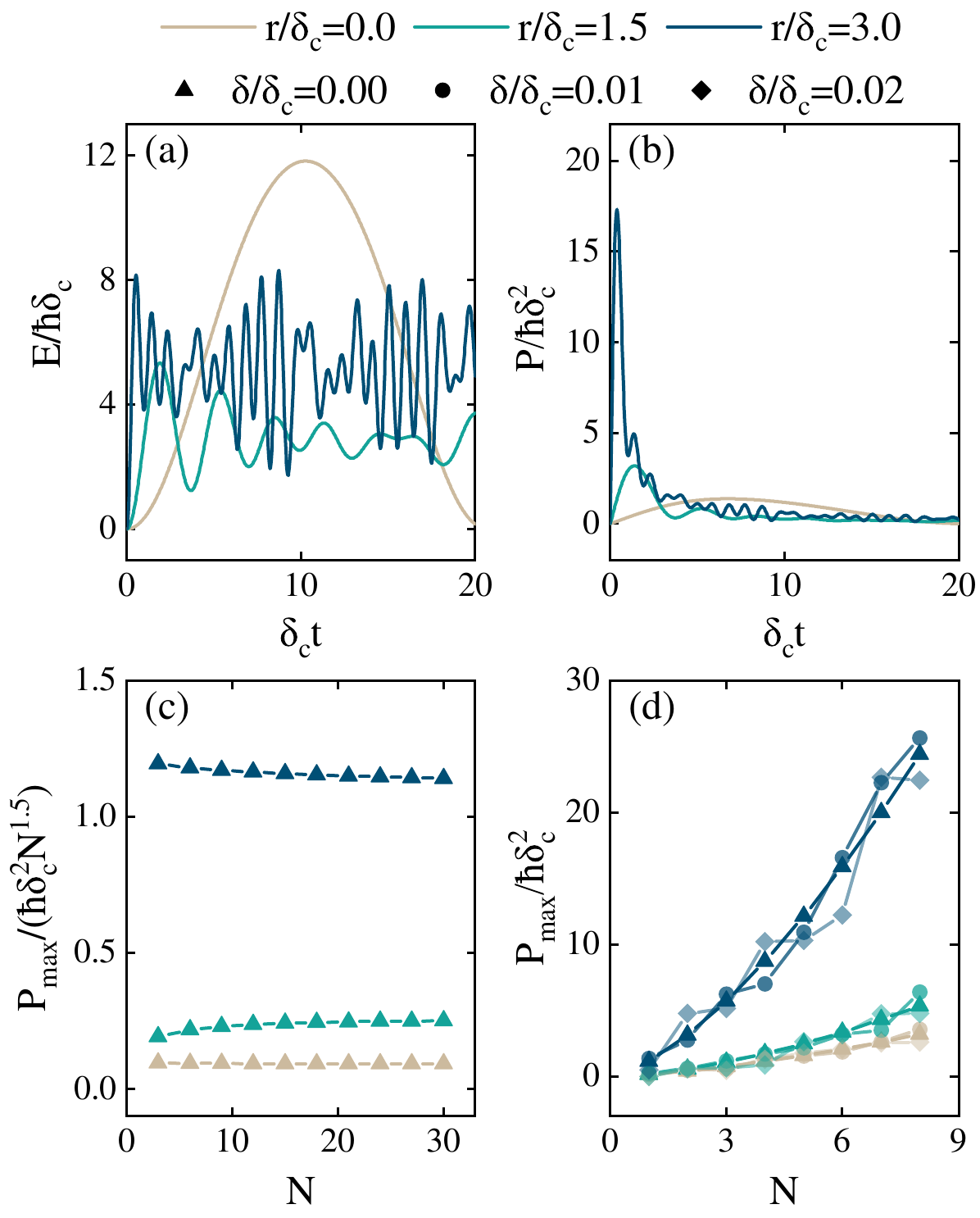}
\caption{Closed-system charging performance. (a) Time evolution of the stored energy and (b) charging power for different squeezing parameters $r/\delta_{c}=0.0$, $1.5$, $3.0$. (c) Scaling of the maximum charging power with the number of qubits for the homogeneous case $\delta/\delta_{c}=0$, normalized by the collective charging advantage $N^{1.5}$ of the standard Dicke model. (d) Same scaling for the inhomogeneous case $\delta/\delta_{c}\neq0$, where static disorder is applied to both qubit frequencies and qubit-cavity coupling strengths. Other parameters are the same as in Fig.~\ref{fig2}.}
\label{fig3}
\end{figure}
In the closed-system scenario without cavity dissipation and qubit decoherence, Figs.~\ref{fig3}(a) and \ref{fig3}(b) show the time evolution of the stored energy and charging power under varying squeezing parameters. The squeezing effect induced by two-photon driving enhances the effective charger-battery coupling but simultaneously makes the counter-rotating terms nonnegligible. These terms break the conservation of total excitation number and introduce virtual-photon-induced quantum fluctuations, preventing the battery from reaching the fully charged state. Nonetheless, the squeezing effect accelerates energy transfer from the charger to the battery, substantially increasing the charging power. Figure~\ref{fig3}(c) displays the scaling of the maximum charging power with the number of qubits. The battery retains the collective charging advantage of the standard Dicke model, with the maximum power following the scaling law $P_{\rm max}=\alpha N^{\beta}$ and the
scaling exponent $\beta=1.5$ \cite{PhysRevLett.120.117702}. Compared to the case without two-photon driving, the coefficient $\alpha$ increases by a factor of 12.5 at a squeezing parameter $r/\delta_{c}=3$, i.e., $P_{\rm max}(r/\delta_{c}=3)/P_{\rm max}(r/\delta_{c}=0)\approx12.5$, and further enhancement can be achieved by increasing either the squeezing parameter or the number of charger modes shown in Appendix~\ref{sectionB}.

In practical implementations, inhomogeneity in qubit frequencies and coupling strengths is unavoidable, especially in superconducting circuits where identical qubits are extremely challenging to realize. To evaluate the experimental feasibility of this scheme, we introduce static disorder drawn from a uniform distribution, setting the frequency of the $j$th qubit as $\omega_{j}=\omega_{q}+\epsilon_{j}$ and its coupling strength to the cavity mode as $g_{j}=g+\eta_{j}$, where $\epsilon_{j}, \eta_{j} \in [-\delta,~\delta]$ are independent random variables. Numerical results presented in Fig.~\ref{fig3}(d) demonstrate that even when the inhomogeneous broadening is on the same order as the bare coupling strength, the exponentially enhanced coupling can effectively sustain the dominance of collective charging effects, allowing the battery to maintain  high-power charging performance with weak sensitivity to parameter perturbations. This robustness not only verifies the strong tolerance of the scheme to non-ideal experimental conditions but also reveals a feasible pathway to overcome inherent parameter dispersion in solid-state systems via quantum control methods.

\begin{figure}[tbp]
\centering
\includegraphics[width=0.45\textwidth]{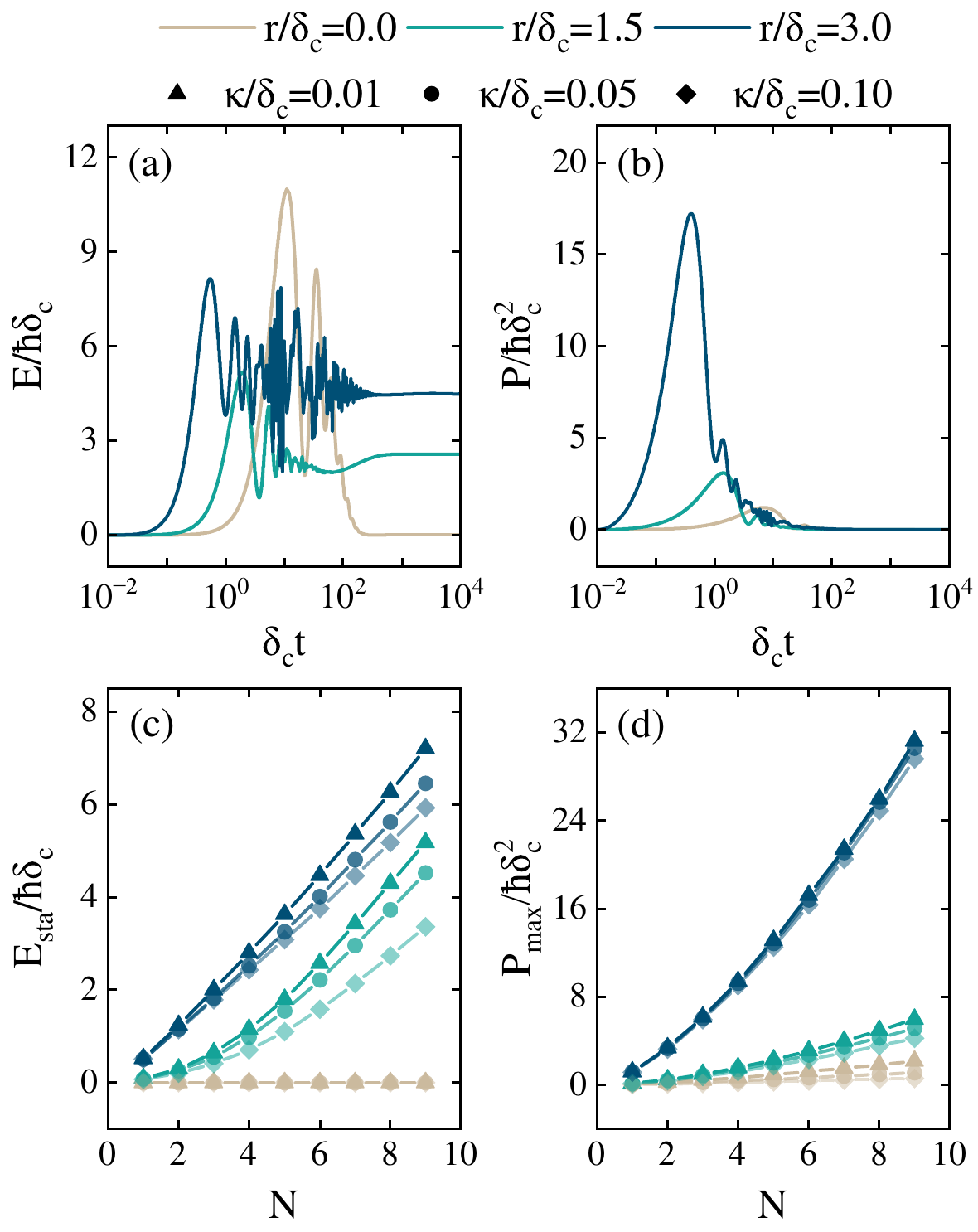}
\caption{Open-system charging performance. (a) Time evolution of the stored energy and (b) charging power for squeezing parameters $r/\delta_{c}=0.0$, $1.5$, $3.0$, with fixed dissipation rates $\kappa/\delta_{c}=\gamma/\delta_{c}=0.01$. (c) Scaling of the steady-state energy and (d) maximum charging power with the number of qubits for dissipation rates $\kappa/\delta_{c}=\gamma/\delta_{c}=0.01$, $0.05$, and $0.10$, representing the underdamped, critically damped, and overdamped regimes, respectively. All other parameters are identical to those in Fig.~\ref{fig2}.}
\label{fig4}
\end{figure}

Under dissipative and decoherence conditions, the battery evolves into a steady state with finite stored energy for $r/\delta_{c} \neq 0$, as shown in Fig.~\ref{fig4}(a). This behavior originates from two synergistic mechanisms: dark-state protection and environmental modification. The dark states of the many-body system satisfy $J_{-}\lvert\text{dark}\rangle=0$, where quantum interference cancels collective dipole radiation, rendering these states immune to decay via the collective relaxation channel $\gamma \mathbb{L}_{J_-,J_+}$ \cite{PhysRevLett.124.013601,Zanner2022,PhysRevApplied.14.024092}. Meanwhile, the squeezing effect modifies the environmental noise shown in Eq.~(\ref{superoperator}), where the $\sinh^2 r \mathbb{L}_{a^\dagger,a}$ term converts vacuum fluctuations into effective thermal excitations, effectively endowing the zero-temperature environment with a non-zero temperature in the squeezed frame. The resulting dynamic balance between energy injection and dissipation drives the battery into a steady state with finite stored energy. As shown in Fig.~\ref{fig4}(b), the charging power retains the collective charging advantage of the closed system, indicating that dark-state protection and squeezing-enhanced coupling jointly preserve many-body cooperative effects. Figures~\ref{fig4}(c) and~\ref{fig4}(d) further demonstrate that even in the critically damped and overdamped regimes, where the dissipation rate approaches or exceeds the coupling strength, the battery sustains a steady state with non-zero stored energy and its associated quantum advantage. 

The impact of two-photon driving exhibits a distinct trade-off depending on whether the system is closed or open. In the closed system, the enhanced charging power comes at the cost of a slight reduction in the maximum stored energy, as counter-rotating terms break excitation number conservation. In the open system, this trade-off yields an advantage: two-photon driving not only maintains high-power charging but also achieves stable energy storage via dark-state protection and squeezing-induced effective thermal excitation. Thus, the two-photon driving scheme exhibits dual superiority: it simultaneously improves charging and storage properties while remaining robust against parameter disorder and environmental noise. This sets a theoretical foundation for the practical application of QBs in imperfect experimental environments.
\section{Conclusions} \label{section4}
In summary, we proposed a superconducting quantum circuit composed of a two-photon-driven LC resonator coupled to an array of transmon qubits, offering new possibilities for the optimization and realization of multipartite QBs. In such system, two-photon driving exponentially enhances the cavity-qubit coupling strength, which in turn induces near-degenerate energy-level structures and highly entangled quantum states. This significantly enhances the charging power and enables rapid energy transfer between the charger and battery. Moreover, the squeezed cavity mode and the quantum correlations reflected in the bimodal Wigner function contribute to coherence protection and decoherence suppression during stable energy storage. We further demonstrated the robustness of the scheme under practical experimental imperfections, including parameter disorder and environmental noise, confirming that the charging and energy storage advantages remain well preserved. This work provides a feasible theoretical framework and an implementation platform for developing high-power, high-stability quantum energy storage devices, and also lays a foundation for the further application of parametric control in quantum information processing and microscopic quantum energy systems.
%\nocite{*}
\section*{Acknowledgments}
The work is supported by the National Natural Science Foundation of China (Grant No.~12475026) and the Natural Science Foundation of Gansu Province (No.~25JRRA799). J.H.A. is supported by the National Natural Science Foundation of China (Grants No. 12275109, No. 92576202, and No. 12247101), the Quantum Science and Technology-National Science and Technology Major Project (Grant No. 2023ZD0300904), the Gansu Science and Technology Leading Talent Program (Grant No. 26RCKA011), and the Fundamental Research Funds for the Central Universities (Grant No. lzujbky-2025-jdzx07). 

\section*{Data availability}
%The data that support the findings of this article are openly available.
The data that support the findings of this article are not publicly available. The data are available from the authors upon reasonable request.
\appendix
\section{Quantum criticality and quantum fluctuation}\label{sectionA} 
Under strong parametric driving, the system considered in this work reduces to the standard Dicke model, enabling the derivation of quantum criticality and quantum fluctuation within that framework \cite{PhysRevE.67.066203,PhysRevLett.124.073602}. The Hamiltonian (\ref{squeezedHamiltonian}) is reformulated as follows
\begin{equation}
H=\widetilde{\delta}_{c}a^{\dag}a+\delta_{q}J_{z}+\frac{g_{e}}{\sqrt{N}}\left(a^{\dag}+a\right)\left(J_{+}+J_{-}\right),
\end{equation}
where the effective coupling strength $g_{e}=-\sqrt{N}\widetilde{g}$. In the thermodynamic limit $N\rightarrow\infty$, the system can be treated via the Holstein-Primakoff transformation, which maps the spin operators onto bosonic ones: $J_{z}=c^{\dag}c-N/2$ and $J_{+}=c^{\dag}\sqrt{N-c^{\dag}c}\approx\sqrt{N}c^{\dag}$. Expanding to leading order in $1/N$ and keeping terms up to quadratic order, the Hamiltonian linearizes to
\begin{equation}
H\approx\widetilde{\delta}_{c}a^{\dag}a+\delta_{q}c^{\dag}c+g_{e}\left(a^{\dag}+a\right)\left(c^{\dag}+c\right).
\end{equation}
The initial state $|\Psi(0)\rangle=|n\rangle\otimes|j, -j\rangle=|n\rangle\otimes|0\rangle$, where $|n\rangle$ denotes a Fock state of the photon mode (with $n=0$ for the vacuum) and $|0\rangle$ is the bosonic vacuum for the qubit ensemble in the Holstein-Primakoff representation.

To determine the critical coupling condition between two bosonic modes, the position and momentum operators are expressed in terms of creation and annihilation operators: $x_{a}=(a+a^{\dag})/\sqrt{2}$, $p_{a}=i(a^{\dag}-a)/\sqrt{2}$, $x_{c}=(c+c^{\dag})/\sqrt{2}$, $p_{c}=i(c^{\dag}-c)/\sqrt{2}$. The Hamiltonian becomes $H=\widetilde{\delta}_{c}\left(p_{a}^{2}+x_{a}^{2}\right)/2+\delta_{q}\left(p_{c}^{2}+x_{c}^{2}\right)/2+2g_{e}x_{a}x_{c}$.
To analyze the stability of the system, the system is cast in matrix form
\begin{equation}
\begin{aligned}
H=&~\frac{1}{2}
\begin{pmatrix}
  p_{a}&p_{c}
\end{pmatrix}
\begin{pmatrix}
  \widetilde{\delta}_{c}&0\\
  0&\delta_{q}
\end{pmatrix}
\begin{pmatrix}
  p_{a}\\p_{c}
\end{pmatrix}\\
&+\frac{1}{2}
\begin{pmatrix}
  x_{a}&x_{c}
\end{pmatrix}
\begin{pmatrix}
  \widetilde{\delta}_{c}&2g_{e}\\
  2g_{e}&\delta_{q}
\end{pmatrix}
\begin{pmatrix}
  x_{a}\\x_{c}
\end{pmatrix},
\end{aligned}
\end{equation}
which separates the contributions from position and momentum operators. The stability of the system requires the coefficient matrix of the potential-energy part to be positive definite, i.e., its determinant must be positive
\begin{equation}
\begin{vmatrix}
  \widetilde{\delta}_{c}&2g_{e}\\
  2g_{e}&\delta_{q}
\end{vmatrix}
=\widetilde{\delta}_{c}\delta_{q}-4g_{e}^{2}>0.
\end{equation}
The onset of instability, which marks the quantum phase transition, occurs when the determinant vanishes. 
Solving for the coupling strength from $\widetilde{\delta}_{c}\delta_{q}=4g_{c}^{2}$ yields the critical value
\begin{equation}
|g_{c}|=\frac{1}{2}\sqrt{\widetilde{\delta}_{c}\delta_{q}}=\frac{1}{2}\sqrt{\delta_{c}\delta_{q}\operatorname{sech}{\left(2r\right)}}.
\end{equation}
The squeezing effect induced by two-photon driving both enhances the effective coupling and decreases the critical coupling, enabling the superradiant phase transition to occur at a significantly reduced coupling strength \cite{PhysRevLett.124.073602}.

In the normal phase $|g_{e}|<|g_{c}|$, the ground state preserves the $\mathbb{Z}_{2}$ symmetry, implying $\langle a\rangle_{G}=0,~\langle c\rangle_{G}=0$. The Hamiltonian expectation and variance in the initial state are evaluated as: $\langle H\rangle=\widetilde{\delta}_{c}n,~\langle H^{2}\rangle=\widetilde{\delta}_{c}^{2}n^{2}+g_{e}^{2}\left(2n+1\right)$, leading to the energy uncertainty
\begin{equation}
\Delta H=\sqrt{\langle H^{2}\rangle-\langle H\rangle^{2}}=\frac{g}{2}{\rm e}^{r}\sqrt{N(2n+1)}.
\end{equation}

In the superradiant phase $|g_{e}|>|g_{c}|$, the system undergoes spontaneous symmetry breaking, and both field and qubit ensemble acquire nonzero expectation values $\langle a\rangle_{G}=\sqrt{N}\alpha_{1},~\langle c\rangle_{G}=\sqrt{N}\beta_{1}$ \cite{PhysRevB.108.235424,PhysRevLett.107.140402}. Introducing displacement $a=d+\sqrt{N}\alpha_{1},~c=e+\sqrt{N}\beta_{1}$, where the displacement parameters are determined as \cite{PhysRevE.67.066203}
\begin{equation}
\alpha_{1}=\frac{g_{e}}{\sqrt{\widetilde{\delta}_{c}\delta_{q}}}\sqrt{1-\frac{g_{c}^{2}}{g_{e}^{2}}},~\beta_{1}=-\frac{1}{2}\sqrt{\frac{\widetilde{\delta}_{c}}{\delta_{q}}}\sqrt{1-\frac{g_{c}^{2}}{g_{e}^{2}}},
\end{equation}
we can obtain an effective quadratic Hamiltonian after removing linear terms and constant shifts
\begin{equation}
H\approx\widetilde{\delta}_{c}d^{\dag}d+\delta_{q}e^{\dag}e+g_{e}\left(d^{\dag}+d\right)\left(e^{\dag}+e\right).
\end{equation}
The initial state becomes a displaced vacuum state
\begin{equation}
|\Psi(0)\rangle=D(-\sqrt{N}\alpha_{1})|n\rangle\otimes D(-\sqrt{N}|\beta_{1}|)|0\rangle.
\end{equation}
%where
%\begin{equation*}
%\begin{aligned}
%&D^{\dag}(-\sqrt{N}\alpha_{1})dD(-\sqrt{N}\alpha_{1})=d-\sqrt{N}\alpha_{1},\\
%&D^{\dag}(-\sqrt{N}|\beta_{1}|)eD(-\sqrt{N}|\beta_{1}|)=e-\sqrt{N}\beta_{1}.
%\end{aligned}
%\end{equation*}
The expectation values of the energy and its square in the initial state are
\begin{equation*}
\begin{aligned}
\langle H\rangle=&~\widetilde{\delta}_{c}n+\widetilde{\delta}_{c}\alpha^{2}N+\delta_{q}\beta^{2}N-4g_{e}\alpha\beta N,\\
\langle H^{2}\rangle=&~\langle H\rangle^{2}+\left[4g_{e}^{2}\left(\alpha^{2}+\beta^{2}\right)-4g_{e}\left(\widetilde{\delta}_{c}+\delta_{q}\right)\alpha\beta\right]N\\
&+\left(2\widetilde{\delta}_{c}^{2}\alpha^{2}+8g_{e}^{2}\beta^{2}-8g_{e}\widetilde{\delta}_{c}\alpha\beta\right)nN\\
&+\left(\widetilde{\delta}_{c}^{2}+2g_{e}^{2}\right)n+g_{e}^{2}.
\end{aligned}
\end{equation*}
In the thermodynamic limit, the energy uncertainty tends to
\begin{equation}
\Delta H\approx\frac{1}{2}\sqrt{\mu{\rm e}^{6r}+\left(\mu+\nu\right){\rm e}^{2r}+2\nu\left(8n+3\right)},
\end{equation}
where $\mu=N^{3}g^{4}/2\delta_{c}\delta_{q}$ and $\nu=N^{2}g^{2}$. Two-photon squeezing enhances quantum fluctuations exponentially due to its inherent nonlinearity, whereas merely increasing the initial photon number yields only linear growth in average energy and square-root growth in energy uncertainty. Hence, two-photon driving is far more efficient at enhancing quantum fluctuations than simply adding more photons.
\section{Charging power bound}\label{sectionB}
Compared to the lumped-parameter LC resonator, the one-dimensional transmission line cavity exhibits distributed characteristics and can be modeled as a cascade of LC circuits \cite{PhysRevA.69.062320}. This architecture allows for precise control over its mode structure, vacuum fluctuations, and coupling strengths through geometric parameters such as characteristic impedance and length \cite{PhysRevA.69.062320,RevModPhys.93.025005}. Under the weak-coupling condition and considering an external drive field propagating along the transmission line, we derive the effective Hamiltonian for a two-photon driven transmission line cavity coupled to transmon qubits. The Hamiltonian reads
\begin{equation}
\begin{aligned}
H=&~\sum_{i=1}^{m}\left[\hbar\omega_{c}a_{i}^{\dag}a_{i}-\hbar\lambda\cos\left(\omega_{p}t\right)\left(a_{i}+a_{i}^{\dag}\right)^{2}\right]\\
&+\hbar\omega_{q}J_{z}-\sum_{i=1}^{m}\hbar g\left(a_{i}J_{+}+a_{i}^{\dag}J_{-}\right),
\end{aligned}
\end{equation}
which consists of $m$ independent cavity modes, the two-photon driving, the free Hamiltonian of $N$ qubits, and the coupling between cavity modes and qubits. To eliminate the explicit time dependence of the drive field and to renormalize the quantum fluctuations of the cavity field quadratures, we introduce two unitary transformations
\begin{equation*}
U_{r}=\prod_{i=1}^{m}{\rm e}^{-i{\omega_{p}t\over2}a_{i}^{\dag}a_{i}}{\rm e}^{-i{\omega_{p}t\over2}J_{z}},~U_{s}=\prod_{i=1}^{m}{\rm e}^{{r\over2}\left(a_{i}^{\dag2}-a_{i}^{2}\right)},
\end{equation*}
and the effective Hamiltonian takes the form
\begin{equation}
\begin{aligned}
\label{multimodeHamiltonian}
H_{s}=&~\sum_{i=1}^{m}\hbar\delta_{c}\operatorname{sech}{\left(2r\right)}a_{i}^{\dag}a_{i}+\hbar\delta_{q}J_{z}\\
&-\sum_{i=1}^{m}\frac{\hbar g}{2}{\rm e}^{r}\left(a_{i}^{\dag}+a_{i}\right)\left(J_{+}+J_{-}\right)\\
&+\sum_{i=1}^{m}\frac{\hbar g}{2}{\rm e}^{-r}\left(a_{i}^{\dag}-a_{i}\right)\left(J_{+}-J_{-}\right).
\end{aligned}
\end{equation}
Here, the system describes a multimode cavity under two-photon driving coupled to an ensemble of qubits. Setting $m=1$ recovers the generalized Dicke model with a correction term discussed in the main text, and further taking $r=0$ reproduces the standard Dicke model.

We then analyze the charging power of the QB defined in this system. The average charging power is defined as the time derivative of the energy in the QB \cite{PhysRevLett.118.150601,PhysRevLett.125.040601}, i.e.,
\begin{equation}
\begin{aligned}
P(t)&={\rm{Tr}}\Big(H_{b}\frac{d\rho_{s}}{dt}\Big)=-i{\rm{Tr}}\left(H_{b}[H_{s},\rho_{s}]\right)\\
&=-i{\rm{Tr}}\left([H_{b},H_{s}]\rho_{s}\right)=-i{\rm{Tr}}\left([H_{b},H_{int}+H_{cor}]\rho_{s}\right),
\end{aligned}
\end{equation}
where $H_{int}$ and $H_{cor}$ correspond to the last two terms of the effective Hamiltonian (\ref{multimodeHamiltonian}). In the derivation, we have used the cyclic property of the trace and the fact that $H_b$ commutes with other terms. Using the operator norm inequality and strong parametric driving, the absolute value of the power satisfies
\begin{equation}
\begin{aligned}
|P(t)|_{r\rightarrow\infty}&=|-i{\rm{Tr}}\left([H_{b},H_{int}]\cdot\rho_{s}\right)|\\
&\leq\|[H_{b},H_{int}]\|\leq2\|H_{b}\|\|H_{int}\|\\
&=\frac{{\rm e}^{r}}{2}|P(t)|_{r=0}=\frac{m{\rm e}^{r}}{2}|P(t)|_{Dicke}.
\end{aligned}
\end{equation}
This inequality reveals that the upper bound of the maximum charging power scales exponentially with the squeezing parameter $r$ and linearly with the number of cavity modes $m$. For $m=1$ and $r=0$, the system reduces to the standard Dicke model, with maximum power $|P(t)|_{Dicke}$ as the benchmark. For $m=1$ and $r>0$, the single-mode squeezed cavity enhances the power bound by a factor of ${\rm e}^{r}/2$ compared to the Dicke model. For $m>1$ and $r>0$, the multimode structure further boosts the power bound by an additional factor of $m$, yielding $(m{\rm e}^{r}/2)|P(t)|_{Dicke}$. Thus, the combination of squeezing and multimode structures synergistically enhances the charging power. This suggests that using multimode transmission line cavities with squeezed light can surpass the power limits of conventional single-mode systems, offering a viable route toward high-performance QBs.
\bibliography{refercence}

% Produces the bibliography via BibTeX.

\end{document}